\documentclass[draft]{agujournal2019}
\usepackage{amsmath}
\usepackage{url} 
\usepackage{lineno}
\usepackage[inline]{trackchanges} 
\usepackage{soul}

\draftfalse

\journalname{Geophysical Research Letters}

\begin{document}

%
%


\title{Transition to double-cell mock Walker circulations with surface warming explained by periodic convection}

%
%




\authors{Heng Quan \affil{1,2}, Yi Zhang \affil{3}, Guy Dagan \affil{4}, Stephan Fueglistaler \affil{1,2}}

\affiliation{1}{Department of Geosciences, Princeton University, Princeton, NJ, US}
\affiliation{2}{Program in Atmospheric and Oceanic Sciences, Princeton University, Princeton, NJ, US}
\affiliation{3}{Courant Institute of Mathematical Sciences, New York University, New York, NY, US}
\affiliation{4}{Institute of Earth Sciences, The Hebrew University of Jerusalem, Jerusalem, Israel}





\correspondingauthor{Heng Quan}{hengquan@princeton.edu}




\begin{keypoints}
\item The transition to double-cell mock Walker circulations with warming is associated with the emergence of periodic convection.

\item Convective and stratiform mode dominate the warm pool alternately in a warmer climate, resulting in the lower and upper cell respectively.



\item The long-lived stratiform mode after the termination of the deep convection is the key to the double-cell circulation.

\end{keypoints}

%
%

%
%



\begin{abstract}
Idealized mock Walker simulations are widely used to study the interactions between overturning circulation and convection in the tropics. Previous studies documented a transition from a single-cell to a double-cell mock Walker circulation when the average sea surface temperature exceeds 300\,K. Here, we ascribe the transition to the emergence of periodic convection with warming due to stronger convectively-coupled waves. In cold simulations, the warm pool is dominated by steady deep convection, which results in a single overturning cell. In hot simulations, the warm pool is alternately dominated by deep convection and a stratiform mode, resulting in lower and upper cells respectively. This study suggests that the Walker circulation in a warmer climate may feature complex structural changes in addition to a weakening in strength, and highlights the profound impacts of convection on overturning circulation.
\end{abstract}


\section*{Plain Language Summary}
The Walker circulation features air rising over the warm western Pacific, flowing eastward high up, sinking over the cooler eastern Pacific, and returning near the surface. Previous studies ran idealized atmospheric model simulations, and found that the single-cell Walker circulation splits into two vertically stacked overturning cells at a critical surface temperature close to today's tropical Pacific average. Here, we trace the transition to how the rising air behaves over the warmest, rainiest region. In a cold simulation, this region is always dominated by air ascending from the surface to the top, which drives a single overturning cell. In a hot simulation, this region is alternately dominated by air ascending from the surface to the top, which drives the lower overturning cell; and air ascending only in the upper level, which drives the upper overturning cell. Together, averaged over time, they form the double-cell structure. Our results suggest that as the climate warms, the tropical atmospheric circulation may not simply weaken but could change shape, implying that localized vertical motion may affect basin-scale overturning circulation.

\section{Introduction}
\label{sec:introduction}

The tropical atmospheric motion has multiple scales ranging from planetary-scale overturning circulation to kilometer-scale convective processes. Multi-scale interactions are important for the tropical climate because they affect circulation patterns, cloud distributions, and their responses to global warming \cite{bony2015}. The mock Walker simulation \cite{raymond1994,kuang2012,wofsy2012} is an idealized tool to study the coupling between overturning circulation and convection. It is a cloud-resolving simulation on a rectangular domain with sea surface temperature (SST) uniform in the short dimension and varying in the long dimension, mimicking the zonal SST gradient in the equatorial Pacific and deriving a Walker-like overturning circulation. The length of the channel ($\mathcal{O}(10000\,\mathrm{km})$) is large enough to resolve tropical-scale circulations, and the resolution ($\mathcal{O}(1\,\mathrm{km})$) is fine enough to partially resolve small-scale convection. The mock Walker simulation has been used to study tropical circulation \cite{grabowski2000,yano2002,kuang2012,quan2025wtg}, the responses of tropical precipitation to global warming \cite{dagan2023convection,sokol2026,quan2026}, and tropical climate feedback \cite{wing2024,quan2025sst} and forcing \cite{dagan2023forcing, sreelekshmi2026}.

\begin{figure}
    \centering
    \includegraphics[width=\linewidth]{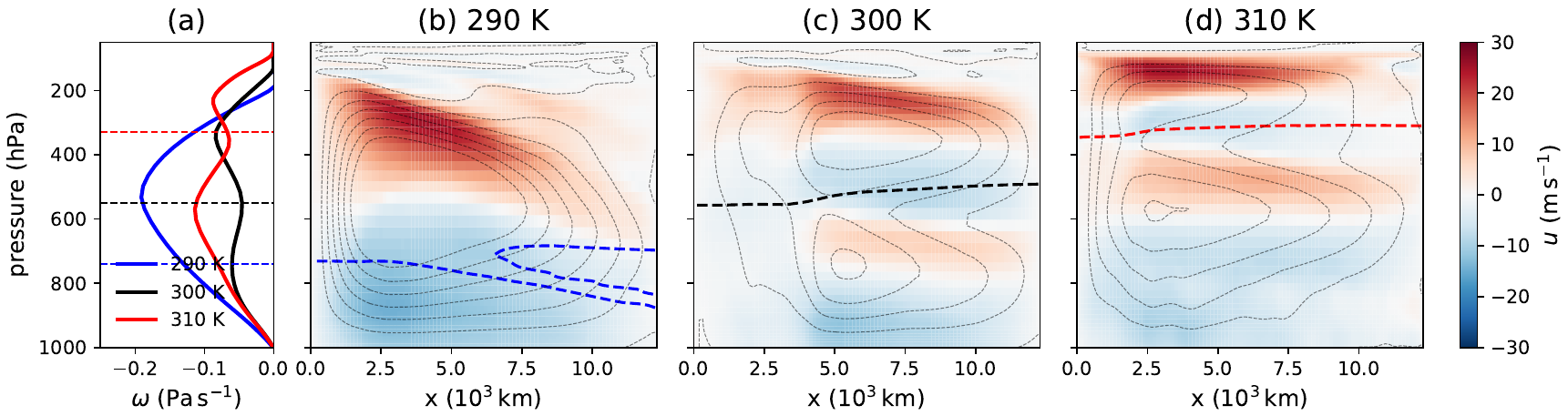}
    \caption{
    (a) The vertical velocity $\omega$ for 3-D SAM mock Walker simulations with mean SSTs of 290\,K (blue), 300\,K (black) and 310\,K (red) averaged over time and over the warm pool (defined as regions with the top 20\% precipitation).
    (b)(c)(d) The time-averaged zonal wind $u$ for 3-D SAM mock Walker simulations (with the prescribed SST linearly decreasing from the left edge to the right edge). The gray contours in the background show the streamfunction $\Psi(x,p) = \int_0^x\omega(x',p) \mathrm{d}x'$ with an interval of 50000\,$\mathrm{kg m^{-1}s^{-1}}$. Solid contours indicate counterclockwise flow and dashed contours indicate clockwise flow.
    The colored dashed lines in all panels are 0$^\circ$C isotherms.
    }
    \label{fig:mock_Walker_uw_tmean}
\end{figure}

Besides the expected weakening of the mean overturning circulation with warming \cite{held2006,vecchi2006}, previous studies using mock Walker simulations documented a transition from a single-cell to a double-cell circulation structure with two vertically stacked overturning cells when the average SST becomes higher \cite{grabowski2000,yano2002,lutsko2024,quan2025wtg} (also see our Figure \ref{fig:mock_Walker_uw_tmean}). This transition affects the distributions of convection and clouds, but the reason for the transition remains puzzling.

A recent study by \citeA{lutsko2024} related the transition to double-cell circulations to the vertical profile of stability over the cold pool. They demonstrated that a mid-tropospheric maximum stability emerges with SST warming for a moist adiabat, which is consistent with a mid-tropospheric minimum descending velocity over the cold pool, as well as a double-cell circulation according to continuity. However, this is not a causal argument. Only the lapse rate over the warm pool is strongly constrained towards a moist adiabatic lapse rate, whereas the lapse rate over the cold pool is not. The free-tropospheric temperature over the cold pool is controlled by the warm pool and is detached from the surface \cite{zhang2020MSE}. Therefore, although intriguing, the mid-tropospheric maximum stability emphasized by \citeA{lutsko2024} cannot fully explain the transition to double-cell mock Walker circulation with warming.

Here, we explain the transition to double-cell mock Walker circulations with warming by the emergence of periodic convection over the warm pool \cite{seeley2021,dagan2023convection,song2024,spaulding2024,quan2026}. A recent study found that precipitation over the warm pool transitions from a steady state to an intense periodic oscillation state as the average SST warms \cite{quan2026}, which is ascribed to the intensification of large-scale convectively-coupled waves. We show that the warm pool is alternately dominated by deep convection (with strong surface precipitation) and a stratiform mode (with little surface precipitation) in hot simulations. Deep convection in the entire troposphere results in the lower overturning cell, while the stratiform mode confined to the upper-troposphere results in the upper overturning cell. By contrast, the warm pool is always dominated by steady deep convection in cold simulations, resulting in a single overturning cell. Our results advance the understanding of tropical circulation structure in a warmer climate, and highlight the importance of the two-way interaction between convection and large-scale dynamics.

The paper is organized as follows. Section \ref{sec:methods} describes the mock Walker simulations used in this study. Section \ref{sec:results} relates the transition to double-cell mock Walker circulations to the emergence of periodic convection over the warm pool, and demonstrates that the periodic convection is necessary for the double-cell. Finally, Section \ref{sec:conclusions} summarizes the conclusions and discusses implications.

\section{Models and simulations}
\label{sec:methods}

We use the System for Atmospheric Modeling (SAM) \cite{khairoutdinov2003} version 6.11.5 cloud-resolving model (CRM) to conduct mock Walker simulations \cite{kuang2012} with prescribed SSTs. The model is nonhydrostatic, uses one-moment bulk microphysics and a simple Smagorinsky-type scheme for subgrid turbulence, and computes the surface sensible heat, latent heat and momentum fluxes based on the Monin–Obukhov similarity theory. All simulations use a $\mathrm{CO_2}$ concentration of 355\,ppm, and the concentrations of other trace gasses are set to the default values. All simulations have a vertical grid of 64 levels, starting at 25\,m and extending up to 27\,km, and the vertical grid spacing increases from 50\,m at the lowest levels to roughly 1\,km at the top of the domain. The model has a rigid lid at the top with a wave-absorbing layer occupying the upper third of the domain to prevent the reflection of gravity waves. The horizontal resolution is 2\,km and the time step is 5\,s. Radiative fluxes are calculated every 5 minutes using the CAM (Community Atmosphere Model) radiation scheme \cite{collins2006}. Following \citeA{lutsko2024}, the incoming solar radiation is fixed at 650.83\,$\mathrm{Wm^{-2}}$. 

The domain size is $12288\,\mathrm{km} \times 128\,\mathrm{km}$ for 3-D mock Walker simulations. We use solid wall boundary conditions at the two edges in the long dimension and periodic boundary conditions in the short dimension. Following \cite{quan2025wtg}, the prescribed SSTs linearly decrease by 8\,K from the left boundary ($x=0$) to the right boundary ($x = 12288$\,km), mimicking the east-west SST gradient across the equatorial Pacific and causing a overturning circulation. The domain average SST ranges from 290\,K to 310\,K with an increment of +5\,K between simulations. Small temperature perturbations are added near the surface at the beginning of the simulation to initialize convection. All simulations are run for 140 days and reach equilibrium after approximately 50 days. All our results are based on the last 40 days of hourly model output and averaged along the short dimension. We define the warm pool as the regions with the top 20\% precipitation, and a different choice (for example, top 10\% precipitation or top 10\% SST) does not affect the results (not shown).

Besides the standard mock Walker simulations described above, we conduct some additional mock Walker simulations with different setups. Simulations not described in the main text are detailed in supplementary Text S1, and a complete list of all simulations is provided in supplementary Table S1.

\section{Results}
\label{sec:results}

\subsection{Transition to double-cell mock Walker circulations withs warming}

Figure \ref{fig:mock_Walker_uw_tmean}(b)(c)(d) reproduces the transition to double-cell mock Walker circulations with surface warming shown in \citeA{lutsko2024}. The time-averaged zonal wind $u$ features a single overturning cell in the 290\,K mock Walker simulation, whereas the 300\,K and 310\,K simulations feature two vertically stacked overturning cells. The SST threshold for the transition, 300\,K, is close to the present average SST in the equatorial Pacific. The double-cell structure expands upwards with SST warming due to the deepening of the troposphere. The upper cell lies between the 600\,hPa level and the 100\,hPa level in the 300\,K simulation, and moves up to above the 400\,hPa level in the 310\,K simulation.

The double-cell zonal wind $u$ is related to a double-peak profile of the vertical velocity ($\omega$) according to the continuity relation. While \citeA{lutsko2024} focused on the subsiding vertical velocity in the cold pool, we instead focus on the ascending vertical velocity in the warm pool. Figure \ref{fig:mock_Walker_uw_tmean}(a) shows $\omega$ averaged over the warm pool (on the left side of the domain) in three simulations. $\omega(p)$ has a single peak corresponding to the single overturning cell in the 290\,K simulation, while it has two peaks corresponding to the lower and upper cells in the 300\,K and 310\,K simulations. The local minimum in $\omega$ between the two peaks is very close to the melting level (i.e., the 0$^\circ$C isotherm), which, as we will show later, is not a coincidence.

The time-averaged circulation structures above are robust to simulation set-ups. Figure S1 in the supporting information shows that the transition to double-cell mock Walker circulations with surface warming also occurs if we replace 3-D simulations with 2-D simulations, if we replace the linear SST profile with a sinusoidal SST profile that has the warm pool in the center, or if we replace the default interactive radiative transfer calculation with a prescribed and fixed radiative cooling rate. In all cases the SST threshold for the transition is around 300\,K, and the local minimum $\omega$ between two peaks, when present, is close to the melting level.

\subsection{Double-cell circulation explained by periodic convection}

\begin{figure}
    \centering
    \includegraphics[width=\linewidth]{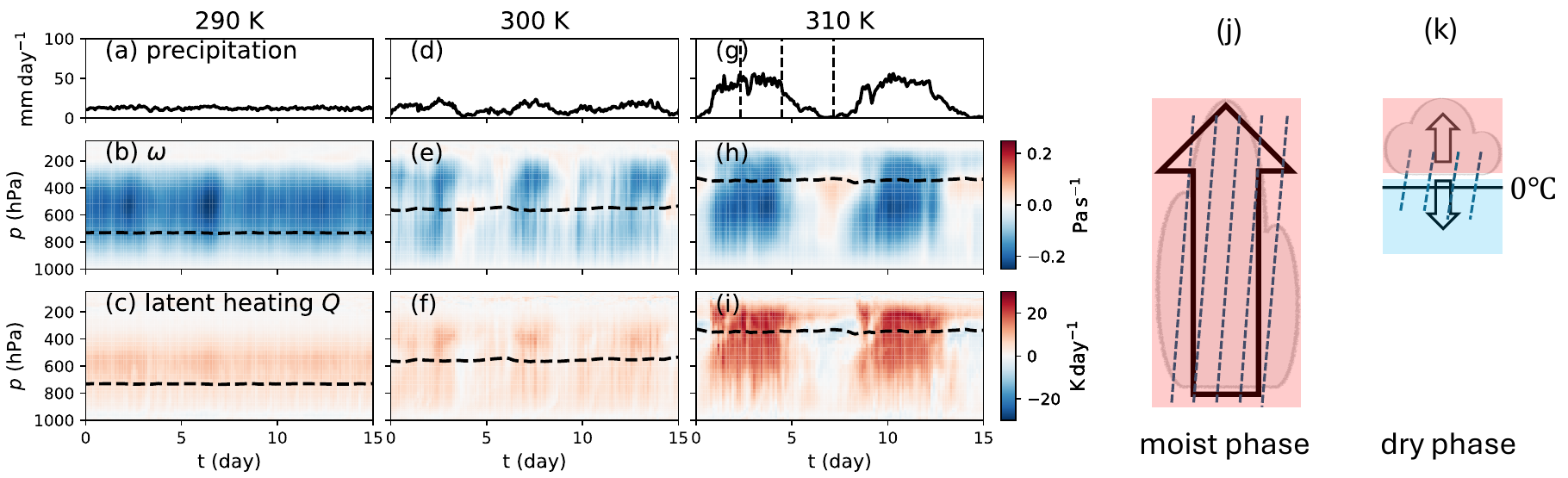}
    \caption{
    (a)(d)(g) The 15-day time series of hourly precipitation averaged over the warm pool for 3-D SAM mock Walker simulations with mean SSTs of 290\,K, 300\,K and 310\,K. The three time slices marked in (g) correspond to three snapshots analyzed in Figure \ref{fig:mock_Walker_3D_310K_tslices}.  
    (b)(e)(h) The corresponding vertical velocity ($\omega$) averaged over the warm pool. 
    (c)(f)(i) The corresponding latent heating rate ($Q$) averaged over the warm pool. In panels b, c, e, f, h and i the dashed lines mark the 0$^\circ$C isotherms.
    (j)(k) Schematics of the moist phases (with high surface precipitation) and the dry phases (with little surface precipitation) over the warm pool of the 310\,K simulation. Arrows represent vertical motion, red shadings represent net latent heating and blue shadings represent net latent cooling.
    }
    \label{fig:mock_Walker_3D_tseries}
\end{figure}

The double-peak $\omega(p)$ profile in Figure \ref{fig:mock_Walker_uw_tmean}(a) co-occurs with a transition from steady convection to periodic convection as the SST warms. Figure \ref{fig:mock_Walker_3D_tseries} shows the time evolution of precipitation, the vertical velocity ($\omega$) and the latent heating rate ($Q$) averaged over the warm pool in a selected 15-day period. The 290\,K simulation (Figure \ref{fig:mock_Walker_3D_tseries}(a)-(c)) features steady deep convection with ascending motion and net latent heating due to condensation throughout the free troposphere. The 300\,K simulation (Figure \ref{fig:mock_Walker_3D_tseries}(d)-(f)) is in an intermediate state with weak oscillatory behavior. The 310\,K simulation (Figure \ref{fig:mock_Walker_3D_tseries}(g)-(i)) features intense periodic oscillation in precipitation, $\omega$ and $Q$. The transition from steady convection to periodic convection with surface warming was documented in \citeA{quan2026} and explained by the intensification of large-scale convectively-coupled waves with warming. This transition is robust to simulation set-ups (Figure S2).

Moist and dry phases of the periodic convection in the warm pool of the 310\,K simulation have distinct vertical structures. During moist phases (for example, day 3) with large surface precipitation, there is ascending motion and net latent heating throughout the free troposphere. As summarized in Figure \ref{fig:mock_Walker_3D_tseries}(j), moist phases are dominated by deep convection. During dry phases (for example, day 7) with little surface precipitation, weaker ascending motion and net latent heating are confined to the upper-troposphere above the 300\,hPa level, while subsidence and net latent cooling dominate the mid-troposphere. As summarized in Figure \ref{fig:mock_Walker_3D_tseries}(k), dry phases are dominated by the stratiform mode \cite{mapes2000,kuang2008instability}, where the condensates aloft hardly fall to the surface due to re-evaporation in the mid-troposphere. These vertical structures in moist and dry phases are unlikely to be artifacts of the microphysical parameterization; they persist when replacing the default single-moment microphysics scheme \cite{khairoutdinov2003} with a double-moment scheme \cite{morrison2005} (Figure S3).

The periodic convection explains the time-averaged double-peak $\omega$ profile in Figure \ref{fig:mock_Walker_uw_tmean}(a) and the double-cell circulation. In the 310\,K simulation, deep convection and the stratiform mode dominate the warm pool alternately (Figure \ref{fig:mock_Walker_3D_tseries}(h)(i)). Deep convection results in the lower peak of the time-averaged $\omega$ in Figure \ref{fig:mock_Walker_uw_tmean}(a) and the lower overturning cell (``lower cell" in this paper refers to the deep circulation with the lower peak in time-averaged $\omega$), while the upper-tropospheric ascending motion of the stratiform mode results in the upper peak of the time-averaged $\omega$ and the upper overturning cell. The subsidence near the melting level of the stratiform mode results in the local minimum time-averaged ascending velocity near the melting level. By contrast, in the 290\,K simulation, the warm pool is always dominated by steady deep convection (Figure \ref{fig:mock_Walker_3D_tseries}(b)(c)), which results in a single peak of the time-averaged $\omega$ in Figure \ref{fig:mock_Walker_uw_tmean}(a) and a single overturning cell.

\begin{figure}
    \centering
    \includegraphics[width=\linewidth]{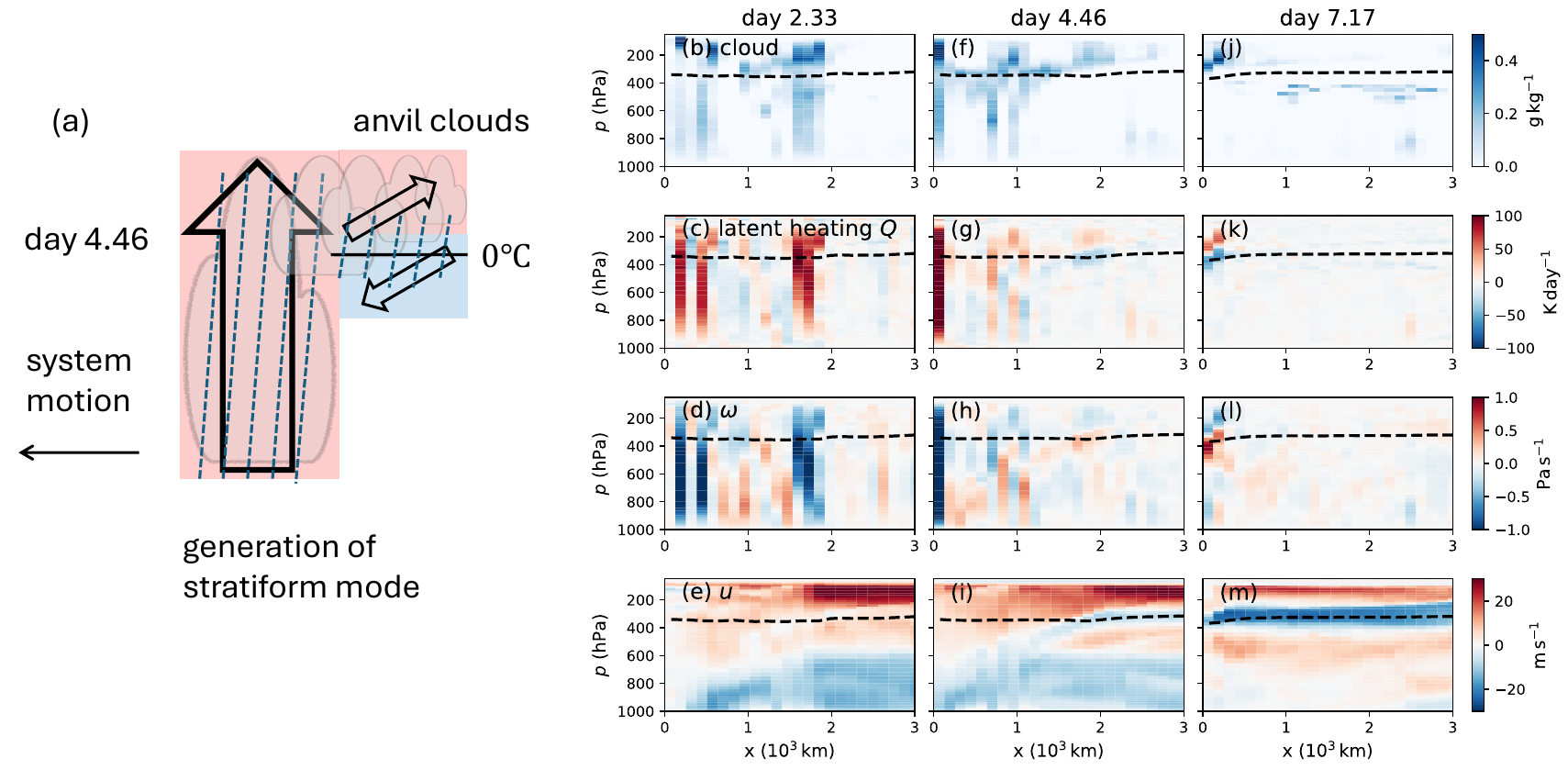}
    \caption{
    (a) A schematic showing the generation of the stratiform region in a mesoscale convective system.  
    (b)-(e) The x-p distribution of the non-precipitating cloud condensates, the latent heating rate ($Q$), the vertical velocity ($\omega$) and the zonal wind ($u$) at day 2.33 over the warm pool of the 310\,K mock Walker simulation. Note that x ranges from 0 to 3000\,km while the length of the domain is 12288\,km.
    (f)-(i) Same as (b)-(e) but at day 4.46.
    (j)-(m) Same as (b)-(e) but at day 7.17.
    The dashed lines in (b)-(m) mark the 0$^\circ$C isotherms.
    }
    \label{fig:mock_Walker_3D_310K_tslices}
\end{figure}

The stratiform mode in our simulations resembles the stratiform regions in mesoscale convective systems (MCSs). As illustrated in Figure \ref{fig:mock_Walker_3D_310K_tslices}(a), the upper-level outflow of the deep convection is directed rearward relative to a moving MCS, creating a front-to-rear slantwise ascending motion \cite{pandya1996} and anvil clouds due to condensation. The region below the cloud base features a rear-to-front inflow as a gravity wave response to latent heating in the convective region \cite{pandya1996}, and this inflow is forced to descend due to the latent cooling of falling precipitation particles \cite{houze2004}. The latent cooling peaks around the melting level where melting, evaporation and sublimation act together \cite{fu2024}, but the subsidence and net latent cooling start above the melting level due to sublimation \cite{gamache1982,braun1997}. A typical MCS has a $\mathcal{O}(100\,\mathrm{km})$ horizontal scale and a $\mathcal{O}(10\,\mathrm{hr})$ lifetime \cite{chen1997}, but the structure in Figure \ref{fig:mock_Walker_3D_310K_tslices}(a) also occurs in large-scale convectively-coupled waves that can span thousands of kilometers and last for several days \cite{kiladis2009}. The $\omega$ and $Q$ averaged over the warm pool in the dry phases of our 310\,K mock Walker simulation (Figure \ref{fig:mock_Walker_3D_tseries}(h)(i)(k)) show a similar dipole structure to the stratiform region in an MCS, with subsidence and net latent cooling starting above the melting level and peaking around the melting level. While the node (where $\omega \approx 0$, $Q \approx 0$) of the stratiform mode is close to the middle of the troposphere in the present climate \cite{mapes2000,kuang2008instability}, it shifts upward in the 310\,K simulation due to a higher melting level (as shown in Figure \ref{fig:mock_Walker_3D_tseries}(k)), so the dipole structure is confined to the upper level.

The snapshots at days 2.33, 4.46 and 7.17 in Figure \ref{fig:mock_Walker_3D_310K_tslices}(b)-(m) confirm that the generation of the stratiform mode in the 310\,K mock Walker simulation is similar to the generation of the stratiform region in MCSs. The three time slices are marked in Figure \ref{fig:mock_Walker_3D_tseries}(g) and are comparable to the growing, mature and dissipating stages in the life cycle of an MCS \cite{robert1982}.
\begin{enumerate}
    \item At day 2.33 (Figure \ref{fig:mock_Walker_3D_310K_tslices}(b)-(e)), there are two deep convective cells at $x \approx 500\,\mathrm{km}$ and $x \approx 1800\,\mathrm{km}$ featuring deep convective clouds, net latent heating, and ascending motion throughout the entire free troposphere. The zonal wind associated with deep convection in panel (e) shows convergence (towards the solid wall) below 600\,hPa and divergence (away from the solid wall) above 600\,hPa, corresponding to the lower overturning cell in Figure \ref{fig:mock_Walker_uw_tmean}(d).
    \item At day 4.46 (Figure \ref{fig:mock_Walker_3D_310K_tslices}(f)-(i)), there is a clear stratiform region ($1000\,\mathrm{km} < x < 2000\,\mathrm{km}$) behind the deep convective cell (Figure S6E in \citeA{quan2026} indicates that convectively-coupled waves propagate towards the warm pool in this simulation.) A front-to-rear slantwise ascending motion creates anvil clouds due to net condensation in the stratiform region, while below the cloud base a rear-to-front inflow is forced to descend due to net latent cooling before reaching the melting level. These structures resemble the generation of the stratiform region in an MCS (Figure \ref{fig:mock_Walker_3D_310K_tslices}(a)). The zonal wind in panel (i) shows both the lower overturning cell (at $x < 1000\,\mathrm{km}$; ``lower cell" in this paper refers to the deep circulation with the lower peak in time-averaged $\omega$) associated with the deep convection and an emerging upper overturning cell (at $x > 1000\,\mathrm{km}$) confined above 400\,hPa associated with the stratiform mode. The vertical shear of the zonal wind (leftward in the lower level, rightward in the upper level) helps to tilt convection and reinforces the generation of the stratiform region.
    \item At day 7.17 (Figure \ref{fig:mock_Walker_3D_310K_tslices}(j)-(m)), the deep convection terminates, but the stratiform mode remains (at $x < 400\,\mathrm{km}$) in the upper troposphere. The upper overturning cell above 400\,hPa associated with the stratiform mode in panel (m) contributes to the time-averaged upper overturning cell in Figure \ref{fig:mock_Walker_uw_tmean}(d).
\end{enumerate}

A video version of Figure \ref{fig:mock_Walker_3D_310K_tslices}(b)-(m) for mock Walker simulations with different SSTs further supports that the periodic convection in the warm pool causes the double-cell overturning circulation (Video S1). In the 290\,K simulation, stratiform regions occasionally emerge behind the deep convection, but there is no clear temporal separation between deep convection and the stratiform mode. Steady deep convection always dominates over the stratiform mode in the warm pool (this is also evident in Figure \ref{fig:mock_Walker_3D_tseries}(b)(c)), so the upper overturning cell associated with the stratiform mode never emerges. By contrast, in the 310\,K simulation, there is clear temporal separation between deep convection and the stratiform mode due to the periodic oscillation. In moist phases, deep convection dominates in the warm pool and creates the lower overturning cell (Figure \ref{fig:mock_Walker_3D_310K_tslices}(e)); in dry phases, the stratiform mode dominates and creates the upper overturning cell (Figure \ref{fig:mock_Walker_3D_310K_tslices}(m)). The stratiform mode alone can persist for 2-3 days during dry phases until deep convection recurs, and this persistence is the key to the time-averaged double-cell circulation.

Before moving on, we briefly comment on the role of ice and the melting level. Because of strong localized latent cooling, the melting level emerges as the level with a local minimum $\omega$ in Figure \ref{fig:mock_Walker_uw_tmean}(a). However, it is not a necessary ingredient for the double-cell circulation. Figure S3 demonstrates that the double-cell circulation still exists in the 310\,K simulation without ice (i.e., water only has a vapor phase and a liquid phase). The theory for the transition to periodic convection with warming \cite{quan2026} invokes large-scale convectively-coupled waves and is not dependent on the presence of ice, so the ice-free 310\,K simulation still features periodic convection in the warm pool (Figure S3), which causes the double-cell circulation.
 
\subsection{Mechanism-denial simulations confirm the necessity of periodic convection for the double-cell circulation}

\begin{figure}
    \centering
    \includegraphics[width=\linewidth]{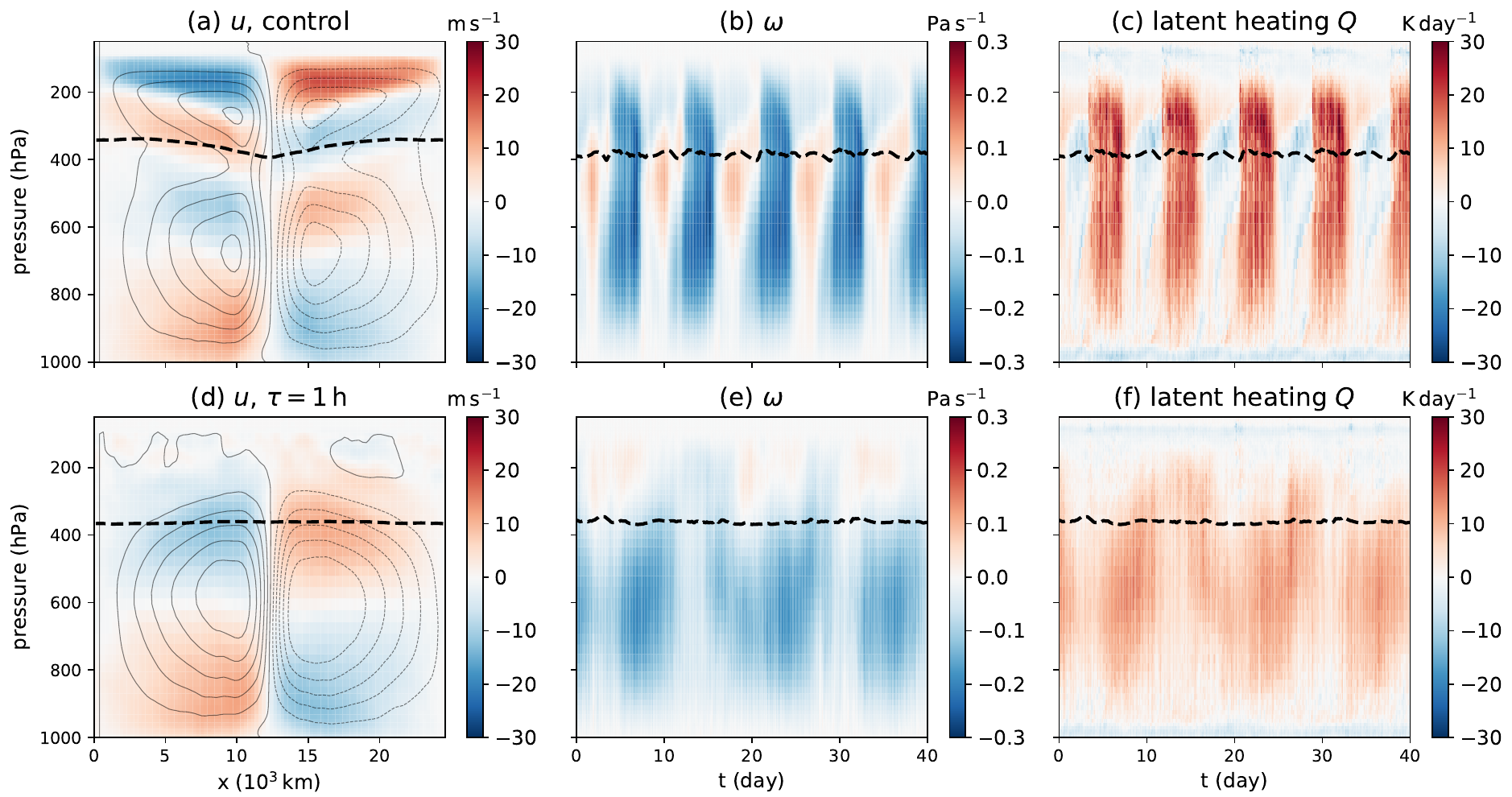}
    \caption{
    (a) The time-averaged zonal wind ($u$) for a 2-D SAM mock Walker control simulation with a sinusoidal SST profile. The domain-averaged SST is 310\,K, and the difference between the warmest SST at the center and the coldest SST at the edges is 5\,K.  The gray contours in the background show the streamfunction $\Psi(x,p) = \int_0^x\omega(x',p) \mathrm{d}x'$ with an interval of 50000\,$\mathrm{kg m^{-1}s^{-1}}$. Solid contours indicate counterclockwise flow and dashed contours indicate clockwise flow. 
    (b)(c) The 40-day evolution of vertical velocity ($\omega$) and latent heating rate ($Q$) averaged over the warm pool of the control simulation.
    (d)(e)(f) Same as (a)(b)(c), but for a mechanism-denial simulation that is identical to the control simulation except for an additional moisture source in the mid-troposphere (details in text).
    The black dashed lines mark the 0$^\circ$C isotherms.
    }
    \label{fig:mock_Walker_qsource_comparison}
\end{figure}

We design a mechanism-denial experiment to demonstrate the necessity of periodic convection for the double-cell circulation. In this case, our control simulation is a 2-D (x-z) mock Walker simulation with a sinusoidal SST profile. Following \citeA{lutsko2024}, the average SST is 310\,K, and the difference between the warmest SST at the center and the coldest SST at the edges is 5\,K. The length of the domain is 24576\,km, and we use a periodic boundary condition. Other settings are identical to Section \ref{sec:methods}. Consistent with the results above, the control simulation features a double-cell overturning circulation (Figure \ref{fig:mock_Walker_qsource_comparison}(a)) due to periodic convection in the warm pool (Figure \ref{fig:mock_Walker_qsource_comparison}(b)(c)), which also shows that deep convection and the stratiform mode dominate the warm pool alternately.

The mechanism-denial simulation features steady deep convection in the warm pool. It is identical to the control simulation, except for an artificial moisture source in the model's moisture budget equation:

\begin{equation}
\label{equ:qsource}
    \frac{\partial q}{\partial t} = \cdots + \mathrm{max}\left\{\frac{0.75q^* - q}{\tau}, 0\right\},
\end{equation}
where $q$ is the specific humidity, $q^*$ is the saturation specific humidity, and $\tau = 1\,\mathrm{h}$ is a moistening timescale. The artificial moisture source is added only when $q/q^*<0.75$ and $263.15\,\mathrm{K} < T < 283.15\,\mathrm{K}$ to keep the mid-tropospheric relative humidity (RH) above 75\%. Intuitively, the artificial moisture source enhances deep convection by moistening the mid-troposphere and reducing the evaporative cooling from entrainment \cite{kuang2008instability,weber2021,ahmed2020}, so we expect that deep convection in the warm pool remains steady and does not decay in the mechanism-denial simulation. In the context of \citeA{quan2026}, the termination of deep convection in the periodic oscillation is related to a positive feedback: An anomalous subsidence perturbation in the mid-troposphere results in a dry anomaly due to vertical moisture advection, which causes an anomalous evaporative cooling that amplifies the subsidence perturbation. The artificial moisture source in the mechanism-denial simulation breaks this positive feedback; hence, it should maintain steady deep convection in the warm pool. Indeed, Figure \ref{fig:mock_Walker_qsource_comparison}(e)(f) shows steady deep convection in the mechanism-denial simulation. Although a weaker oscillatory behavior still exists, the warm pool is always dominated by deep convection, while the stratiform-only mode never emerges.

With steady deep convection in the warm pool, the mechanism-denial simulation has a single-cell overturning circulation instead of a double-cell structure (Figure \ref{fig:mock_Walker_qsource_comparison}(d)). The lower cell associated with deep convection is similar to that in the control simulation, while the upper cell associated with the stratiform mode disappears. Therefore, the mechanism-denial simulation verifies that periodic convection is necessary for the double-cell circulation.

\section{Conclusions and discussions}
\label{sec:conclusions}

We study the transition from single-cell to double-cell mock Walker circulations with SST warming, and ascribe it to the emergence of periodic convection in the warm pool discussed in our preceding paper \cite{quan2026}. In a cold simulation, the warm pool is always dominated by steady deep convection, which results in a single overturning cell. In a hot simulation, the warm pool is alternately dominated by deep convection (in moist phases) and the stratiform mode (in dry phases). They create the lower and upper cells respectively, and result in a time-averaged double-cell circulation. We show that the generation of the stratiform mode is similar to the generation of the stratiform region in a moving MCS, and the persistence of the stratiform mode without deep convection is the key to the double-cell circulation. Our mechanism-denial simulations with periodic convection disabled further confirm that periodic convection is necessary for the double-cell circulation.

One limitation of this study is that we only use a single idealized cloud-resolving model. We also analyzed the output of mock Walker simulations based on other cloud-resolving models in RCEMIP-II \cite{wing2024}, but most of them have an SST contrast of 1.25\,K between the warm pool and the cold pool, which is not large enough to generate a clear overturning circulation as shown in our Figure \ref{fig:mock_Walker_uw_tmean}(b)-(d) (results not shown). Therefore, mock Walker simulations based on other models and with larger ($>5$\,K) SST contrasts can help verify the theories in this study. Meanwhile, while the SST threshold (near 300\,K) for the transition to double-cell mock Walker circulations is close to the present-day average SST in the equatorial Pacific, there is no double-cell circulation in the reanalysis data \cite{lutsko2024}. \citeA{quan2026} found that convectively-coupled Kelvin waves intensify with warming and trigger periodic convection in the western Pacific warm pool in more realistic global simulations. Convectively-coupled gravity waves in our 310\,K mock Walker simulation propagate towards the warm pool (Figure S6E in \citeA{quan2026}), while convectively-coupled Kelvin waves in the real world propagate eastward and away from the warm pool in the equatorial Pacific. The direction of the zonal velocity associated with the stratiform mode in Figure \ref{fig:mock_Walker_3D_310K_tslices}(a) depends on the direction of propagating convectively-coupled waves, but the vertical velocity does not. Therefore, the double-peak vertical velocity profile in Figure \ref{fig:mock_Walker_uw_tmean}(a) may also emerge with warming in the real world. Future work may study the responses of the real-world Walker circulation to global warming in global simulations, and check whether a double-cell Walker circulation emerges together with periodic convection in the warm pool. 

This study suggests that the Walker circulation in a warmer climate may feature complex structural changes in addition to a weakening of strength, and highlights the importance of the two-way interaction between convection and large-scale dynamics.

\section{Open Research}
The source code for SAM can be found at \url{http://rossby.msrc.sunysb.edu/SAM.html}. The source code for SAM with the artificial moisture source (equation \ref{equ:qsource}) can be found at \url{https://zenodo.org/records/12646430}.







\acknowledgments
The authors gratefully acknowledge Marat F. Khairoutdinov for creating and maintaining SAM and Nicholas Lutsko for developing the method that adds the artificial moisture source to SAM.


\bibliography{agusample}

\end{document}